\documentclass[preprint,prb]{revtex4}
\newcommand{\h}{\hspace*{4ex}}

\newcommand \be{\begin{equation}}
\newcommand \ee{\end{equation}}

\begin{document}

\title{\textbf{Quantum Dynamics of Molecular Nanomagnets in a Resonant
Cavity and the Maser Effect}}
\author{C.A. Dartora$^{1,}$\footnote{%
cadartora@eletrica.ufpr.br} \ and \ G.G. Cabrera$^{2,}$\footnote{%
cabrera@ifi.unicamp.br}}
\affiliation{$^1$Electrical Engineering Department, Federal University of
Parana (UFPR),\\
$^2$ Institute of Physics Gleb Wataghin, State University of Campinas
(UNICAMP)\\
}

\begin{abstract}
{We study the dynamics of molecular nanomagnets through a fully quantum
mechanical model describing high-spin and high-anisotropy magnetic molecules
subjected to a time-dependent magnetic field along the quantization axis,
which continuously inverts the population of spin states. Crystals of
molecular nanomagnets placed inside a resonant cavity interact with a
quantized electromagnetic field. Relaxation of excited states takes place by
means of spin-photon interaction, allowing stimulated emission of radiation
and creating a maser effect.}\\

\end{abstract}
\maketitle

\section{Introduction}

\h Recent interest on high-spin high-anisotropy molecular
nanomagnets has grown significantly due to the spectacular magnetic effects
they exhibit, namely the pronounced magnetic hysteresis and the quantization
of the magnetic moment\cite{[1]}. Potential applications of such magnets for
current and future technologies include information storage, construction of
nanomagnetic Maser-like devices, and quantum computation. The name
Single-Molecule Magnets (SSM) has been coined to mean that individual
molecules act as magnets. They can be prepared in long-lived excited quantum
states by simply applying a magnetic field\cite{[2]}, and exhibit a stepwise
magnetic hysteresis in a time-dependent magnetic field. Quantum relaxation
of spin states at low temperatures is very sluggish, and has been
intensively studied within the framework of the
Landau-Zener-Stueckelberg(LZS) effect\cite{[3],[4],[5]}.

A detailed study of quantum spin transitions and the effect of the
environment in an experiment with a rotating magnetic field, was done in
Ref. \onlinecite{[6]}. Their approach follows the lines of the LZS theory, taking as the
starting point the following hamiltonian\cite{[6],[7],[8]}: 
\begin{equation}
\mathcal{H}=-DS_{z}^{2}-g\mu _{B}\mathbf{B}\cdot \mathbf{S},  \label{Ham1}
\end{equation}%
where $D$ is the uniaxial anisotropy constant, $g$ is the gyromagnetic
factor, $\mu _{B}$ is the Bohr magneton, $\mathbf{B}$ is the magnetic field
and $S$ is the molecular spin in units of $\hbar $. In the absence of an
applied field $\mathbf{B}$, the ground state is double degenerate,
corresponding to the states with parallel ($m=S$) and anti-parallel ($m=-S$)
projections of the magnetic moment along the quantization axis. Throughout
this paper, we will consider only these two states ($m=-S$ and $m=+S$) as
relevant to the problem\cite{[6]}. In this context, a striking effect is predicted\cite{[9]}, 
where molecular nanomagnetic crystals exhibit a giant magnetic
relaxation due to Dicke superradiance of electromagnetic waves \cite{[dicke]}. In recent
electron spin resonance (EPR) experiments, it has been observed a
pronounceable resonant absorption of electromagnetic radiation by molecular
nanomagnets\cite{[10],[11],[12],[13],[14]}. In turn, they become a powerful source of coherent
electromagnetic radiation when the wavelength of the emitted photons exceeds
the linear size of crystals. When this condition is achieved, the molecules
can coherently interact with the radiation emmitted, and the phase of the
emitted photons may be considered the same throughout the sample\cite{[9]}. Inside
a resonant cavity, molecular magnets exhibit a strong dependence of the
magnetization on the geometry of the cavity and this effect was observed
experimentally, providing strong evidence for the coherent microwave
radiation given off by the crystals. These observations open the possibility
of building nanomagnetic microwave lasers pumped by magnetic fields\cite{[15]}.

Our aim in this paper is to study the effect of stimulated radiation from
molecular nanomagnetic crystals, by considering a fully quantum mechanical
model, in which the electromagnetic radiation enters as a quantized field.
The essential ingredients in our model include:

\begin{enumerate}
\item[\emph{i)}] hamiltonian (\ref{Ham1}) describing individual molecules,
with a time-varying magnetic field $B_z(t)$ along the quantization axis.
This way, the low and high energy states will be continuously changing with
time, creating the effect of spin-states population inversion. Note that
this is different from most models proposed in the literature, where the
magnetic field $B_z$ is constant, and the time-dependent field is applied in
a transverse direction;

\item[\emph{ii)}] a quantized electromagnetic field inside a resonant
cavity, which allows the relaxation of excited states by means of coherent
photon emission, enhancing the radiation field.
\end{enumerate}

Finally, we will analize the semiclassical limit, in which the
photon-field can be treated as a classical electromagnetic field.

The above program will be developed in the present contribution. The content
of this paper can be described as follows: in the next Section, we formulate
the theoretical basis for analyzing spin dynamics, discussing the
Hamiltonian. In Section III, we analyze the dynamics of a pure quantum
mechanical state initially prepared in one of the double degenerate ground
states of the system. We compute the correlation amplitude between the
initial and the evolved state. Analytical approximated expressions are
obtained to be compared with numerical results. In Section IV, we consider
the photon field as a classical variable and obtain, in close analytical
way, the conditions for a maser-like effect. Finally, in the last Section a
few conclusions and remarks are added.

\section{Hamiltonian and Quantum Dynamics}

\hspace*{4ex} We start considering the two lowest-level states ($m=+S$ and $%
m=-S$) of spin Hamiltonian (\ref{Ham1}) and a second quantized term in
photon variables describing the radiation field in the cavity: 
\begin{equation}
\mathcal{H}_{0}=-DS^{2}\left( 
\begin{array}{cc}
1 & 0 \\ 
0 & 1%
\end{array}%
\right) -g\mu _{B}SB_{0}(t)\left( 
\begin{array}{cc}
1 & 0 \\ 
0 & -1%
\end{array}%
\right) +\hbar \omega a^{\dagger }a\,\left( 
\begin{array}{cc}
1 & 0 \\ 
0 & 1%
\end{array}%
\right) \,,  \label{H1}
\end{equation}%
where the Pauli matrix%
\[
\mathbf{\sigma }_{z}=\left( 
\begin{array}{cc}
1 & 0 \\ 
0 & -1%
\end{array}%
\right) 
\]%
yields the splitting of the levels in the presence of a time-dependent
magnetic field $B_{0}(t)$ applied along the quantization axis. The operator $%
a^{\dagger }(a)$ is the photon creation (annihilation) operator. Inside a
resonant cavity, only photons with a preselected frequency $\omega $ can be
emitted or absorbed. The basis of $\mathcal{H}_{0}$ is given \ through the
kets $\{|S,n>\}$ and $\{|-S,n>\}$, being $n$ the number of photons in a
given state. We will consider that a general state can be written as
follows: 
\[
|\Psi >=\{A_{S}|S>+A_{-S}|-S>\}\otimes \sum_{n=0}^{\infty }\alpha _{n}|n> 
\]

The interaction between the molecular spin and the dipolar component of the
electromagnetic radiation inside the cavity is considered as a perturbation: 
\begin{equation}
\mathcal{H}_{I}=-\hbar \Gamma \left( 
\begin{array}{cc}
0 & 1 \\ 
1 & 0%
\end{array}%
\right) (a^{\dagger }+a)\ ,  \label{H2}
\end{equation}%
where the Pauli matrix 
\[
\mathbf{\sigma }_{x}=\left( 
\begin{array}{cc}
0 & 1 \\ 
1 & 0%
\end{array}%
\right) 
\]%
flip the molecule from one level to the other (with the emission or
absorption of a photon). One can see that our model consists of a two level
system coupled to a quantized harmonic oscilator. Despite the simplicity of
the above mentioned model, no exact analytic solution is yet known\cite{[16]}.
Now, we consider the quantum dynamics in the interaction picture, taking
into account that the non-perturbed (diagonal) hamiltonian $\mathcal{H}_{0}$
is time dependent. In this case, operators evolve according to the equation: 
\[
i\hbar \frac{\partial \mathcal{O}}{\partial t}=[\mathcal{O},\mathcal{H}_{0}] 
\]%
and for the interaction Hamiltonian we get 
\begin{equation}
\mathcal{H}_{I}(t)=e^{iW(t)}\mathcal{H}_{I}e^{-iW(t)}\ ,  \label{Hievolv}
\end{equation}%
being 
\begin{equation}
W(t)=\frac{1}{\hbar }\int_{0}^{t}\mathcal{H}_{0}(t^{\prime })dt^{\prime }
\label{Wt}
\end{equation}%
The temporal evolution of an initial state ket $|\Psi _{0}>$ will be given
by the unitary evolution operator $U_{I}$ in the interaction picture. The
latter can be written in the form of a Dyson series: 
\begin{eqnarray}
U_{I}(t,0) &=&1-\frac{i}{\hbar }\int_{0}^{t}dt_{1}e^{iW(t_{1})}\mathcal{H}%
_{I}e^{-iW(t_{1})}+  \nonumber \\
&&+\left( \frac{-i}{\hbar }\right) ^{2}\int_{0}^{t}dt_{1}e^{iW(t_{1})}%
\mathcal{H}_{I}e^{-iW(t_{1})}\int_{0}^{t_{1}}dt_{2}e^{iW(t_{2})}\mathcal{H}%
_{I}e^{-iW(t_{2})}+...\ ,  \label{UI}
\end{eqnarray}%
with $|\Psi (t)>=U_{I}(t,0)|\Psi _{0}>$. We can express the interaction
hamiltonian $\mathcal{H}_{I}$ in the form below: 
\begin{equation}
e^{iW(t)}\mathcal{H}_{I}e^{-iW(t)}=-\hbar \Gamma \left( 
\begin{array}{cc}
0 & H_{12} \\ 
H_{21} & 0%
\end{array}%
\right) ,  \label{EqHI}
\end{equation}%
being 
\begin{eqnarray}
H_{12} &=&a^{\dagger }\exp \left[ i\left( \omega t-\frac{2g\mu _{B}S}{\hbar }%
\int_{0}^{t}B_{0}(t^{\prime })dt^{\prime }\right) \right] +  \nonumber \\
&&+a\exp \left[ -i\left( \omega t+\frac{2g\mu _{B}S}{\hbar }%
\int_{0}^{t}B_{0}(t^{\prime })dt^{\prime }\right) \right]  \label{H12a} \\
H_{21} &=&H_{12}^{\dagger }=a^{\dagger }\exp \left[ i\left( \omega t+\frac{%
2g\mu _{B}S}{\hbar }\int_{0}^{t}B_{0}(t^{\prime })dt^{\prime }\right) \right]
+  \nonumber \\
&&+a\exp \left[ -i\left( \omega t-\frac{2g\mu _{B}S}{\hbar }%
\int_{0}^{t}B_{0}(t^{\prime })dt^{\prime }\right) \right]  \label{H21a}
\end{eqnarray}%
One must remember that $H_{12}$ and $H_{21}$ are infinite-dimensional in the
Fock space of photons. We write $U_{I}$ in the following form 
\begin{eqnarray}
U_{I} &=&1+i\Gamma \int_{0}^{t}dt_{1}\left( 
\begin{array}{cc}
0 & H_{12}(t_{1}) \\ 
H_{21}(t_{1}) & 0%
\end{array}%
\right) -  \nonumber \\
&&-\Gamma ^{2}\int_{0}^{t}dt_{1}\int_{0}^{t_{1}}dt_{2}\left( 
\begin{array}{cc}
H_{12}(t_{1})H_{21}(t_{2}) & 0 \\ 
0 & H_{21}(t_{1})H_{12}(t_{2})%
\end{array}%
\right) +...  \label{Evolution}
\end{eqnarray}%
The above expressions admit any time-varying magnetic field $B_{0}(t)$. For
sake of convenience, we will restrict our attention to the sinusoidal
dependence: 
\begin{equation}
B_{0}(t)=B_{0}\cos (\Omega t),  \label{bot}
\end{equation}%
and in this case (\ref{H12a}) and (\ref{H21a}) reduce to 
\begin{eqnarray}
H_{12}(t) &=&a^{\dagger }\exp \left[ i\left( \omega t-r\sin (\Omega
t)\right) \right] +a\exp \left[ -i\left( \omega t+r\sin (\Omega t)\right) %
\right] ,  \label{H12b} \\
H_{21}(t) &=&a^{\dagger }\exp \left[ i\left( \omega t+r\sin (\Omega
t)\right) \right] +a\exp \left[ -i\left( \omega t-r\sin (\Omega t)\right) %
\right] ,  \label{H21b}
\end{eqnarray}%
where we have defined 
\begin{equation}
r=\frac{2g\mu _{B}SB_{0}}{\hbar \Omega }.  \label{rdef}
\end{equation}%
The frequency $\Gamma $ in $\mathcal{H}_{I}$ represents the coupling with
the dipolar field, and is small when the wavelength of the radiation field
is far longer than the molecular dimension. In this case, first order
perturbation theory applies, leading to: 
\begin{equation}
U_{I}=1+i\Gamma \mathcal{M}_{1}+O(\Gamma ^{2})  \label{Uifirstorder}
\end{equation}%
and 
\begin{equation}
\mathcal{M}_{1}=\left( 
\begin{array}{cc}
0 & a^{\dagger }F_{1}+aF_{2} \\ 
a^{\dagger }F_{2}^{\ast }+aF_{1}^{\ast } & 0%
\end{array}%
\right) .  \label{defM1}
\end{equation}%
The dynamics of the system will be governed by the functions $F_{1}(t)$ and $%
F_{2}(t)$: 
\begin{eqnarray}
F_{1}(t) &=&\int_{0}^{t}dt_{1}\exp [i[\omega t_{1}-r\sin (\Omega t_{1})]]=%
\frac{J_{0}(r)}{i\omega }(e^{i\omega t}-1)+  \nonumber \\
&&+2\sum_{m=0}^{\infty }\left[ \frac{J_{2m+2}(r)}{\omega
^{2}-(2m+2)^{2}\Omega ^{2}}\Lambda _{1,m}(t)-i\frac{J_{2m+1}(r)}{\omega
^{2}-(2m+1)^{2}\Omega ^{2}}\Lambda _{2,m}(t)\right] ,  \label{F1} \\
F_{2}(t) &=&\int_{0}^{t}dt_{1}\exp [-i[\omega t_{1}+r\sin (\Omega t_{1})]]=%
\frac{J_{0}(r)}{i\omega }(1-e^{-i\omega t})+  \nonumber \\
&&+2\sum_{m=0}^{\infty }\left[ \frac{J_{2m+2}(r)}{\omega
^{2}-(2m+2)^{2}\Omega ^{2}}\Lambda _{1,m}^{\ast }(t)-i\frac{J_{2m+1}(r)}{%
\omega ^{2}-(2m+1)^{2}\Omega ^{2}}\Lambda _{2,m}^{\ast }(t)\right] ,
\label{F2}
\end{eqnarray}%
being $J_{2m+1}(r)$ and $J_{2m+2}(r)$ the Bessel functions of first kind,
and 
\begin{eqnarray}
\Lambda _{1,m}(t) &=&i\omega -i\omega e^{i\omega t}\cos [(2m+2)\Omega
t]-(2m+2)\Omega e^{i\omega t}\sin [(2m+2)\Omega t]  \label{lambda1} \\
\Lambda _{2,m}(t) &=&-i\omega e^{i\omega t}\sin [(2m+1)\Omega
t]+(2m+1)\Omega e^{i\omega t}\cos [(2m+1)\Omega t]-(2m+1)\Omega .
\label{lambda2}
\end{eqnarray}%
>From the above formulae, a resonant behavior occurs when the photon
frequency $\omega $ is an integer multiple of $\Omega $, say $\omega
=m\Omega ,$ and the dominant behavior of $F_{1}$ and $F_{2}$ near that
condition is given by: 
\begin{eqnarray}
F_{1} &\approx &\frac{J_{m}(r)[(-1)^{m}i(1-e^{2i\omega t})+2t\omega ]}{%
2\omega }  \label{F1as} \\
F_{2} &\approx &(-1)^{m}F_{1}^{\ast }=(-1)^{m}\frac{%
J_{m}(r)[(-1)^{m+1}i(1-e^{-2i\omega t})+2t\omega ]}{2\omega }  \label{F2as}
\end{eqnarray}%
In the next Section we will analyze the dynamics of a pure quantum
mechanical state initially prepared as an eigenstate of $\mathcal{H}_{0}$.

\section{Dynamics of Quantum States}

\hspace*{4ex} To fix ideas, consider a quantum state prepared in the
magnetic state $|S>$ of the molecule, in the form: 
\begin{equation}
|\Psi _{0}>=|S>\sum_{n=0}^{\infty }\alpha _{n}|n>  \label{psi0}
\end{equation}%
where the mean number of photons at $\ t=0$ is given by 
\[
n_{0}=<a^{\dagger }a>_{0}=\sum_{n=0}^{\infty }n|\alpha _{n}|^{2}. 
\]%
To first order \ perturbation theory, the evolved state $|\Psi
(t)>=U_{I}|\Psi _{0}>$ reads 
\begin{eqnarray}
|\Psi (t) &>&\approx \frac{1}{\sqrt{1+\Gamma
^{2}[(n_{0}+1)|F_{2}|^{2}+n_{0}|F_{1}|^{2}}}\times  \nonumber \\
&&\times \left[ |\Psi _{0}>+i\Gamma |-S>\sum_{n=0}^{\infty }\alpha
_{n}\left( F_{2}^{\ast }\sqrt{n+1}|n+1>+F_{1}^{\ast }\sqrt{n}|n-1>\right) %
\right] .  \label{psievolved}
\end{eqnarray}%
Observe that we have renormalized the evolved state. The correlation
probability function defined as $C(t)=|<\Psi _{0}|\Psi (t)>|^{2}$, which is
simply the probability of finding the system in the initial state at later
times, yields 
\begin{equation}
C(t)=\frac{1}{1+\Gamma ^{2}[(n_{0}+1)|F_{2}|^{2}+n_{0}|F_{1}|^{2}]}\ ,
\label{Cdet}
\end{equation}%
and the above expression can be approximated at the resonant condition $%
\omega =m\Omega $ by the behavior of $F_{1}$ and $F_{2}$ given in (\ref{F1as}%
) and (\ref{F2as}): 
\begin{equation}
C(t)\approx \frac{1}{1+\Gamma ^{2}|F_{1}|^{2}(2n_{0}+1)},  \label{Cor1}
\end{equation}%
with 
\[
|F_{1}|^{2}=\frac{|J_{m}(r)|^{2}}{4\omega ^{2}}\left( 4\sin ^{2}\omega
t+4t\omega \sin {2\omega t}+4\omega ^{2}t^{2}\right) . 
\]%
For long times $(t\omega \gg 1)$, the leading behavior is quadratic in time%
\[
|F_{1}|^{2}\approx |J_{m}(r)|^{2}t^{2}\ , 
\]%
and we can approximate the above expression as follows: 
\begin{equation}
C(t)\approx \frac{1}{1+t^{2}/\tau ^{2}}\ ,  \label{cor2}
\end{equation}%
with the correlation time $\tau $ defined as: 
\begin{equation}
\frac{1}{\tau }=\Gamma |J_{m}(r)|\sqrt{2n_{0}+1}.  \label{corrTime}
\end{equation}%
The value $1/\tau $ can be interpreted as the rate of emission of photons,
since the nanomagnet will relax to lower energy states by emitting photons.
For large values of the constant $r$, maxima of the transition rate will be
given by $r=2gS\mu _{B}B_{0}/(\hbar \omega /m)\sim \zeta \pi $, with $\zeta
\gg 1$ a constant depending on the order of the Bessel function, yielding 
\[
\left( \frac{1}{\tau }\right) _{\max }\approx \Gamma \sqrt{\frac{\hbar
(\omega /m))}{\pi gS\mu _{B}B_{0}}}\sqrt{2n_{0}+1}. 
\]

Next, we show some examples. To illustrate the evolution of a quantum state,
we consider the initial mean number of photons $n_{0}=0$, \emph{i.e. }the
vacuum for the cavity field, and the molecular spin in the $S$ state (in
this case the initial state is $|\Psi _{0}>=|S,0>$). The resonant cavity is
adjusted to the frequency $f=\omega /2\pi =10$ GHz and the spin-photon
interaction constant is taken as $\Gamma =2$ GHz. In Fig. 1 to 3, we show
the correlation probability $C(t)$ as a function of time, for given values
of $\ r$ and $m=\omega /\Omega $. At a given ratio $m=\omega /\Omega ,$ the
parameter $r$ was set to give the maximum value of the Bessel function $%
J_{m}(r)$. The parameters for the first two figures satisfy the resonant
condition ($m$ an integer). One can see that the behavior for $m$ even or
odd is qualitatively the same. Except for the stepwise character of the
correlation function, $C(t)$ can be estimated by the expression (\ref{cor2}%
), with the relaxation time given by (\ref{corrTime}). Note the irreversible
character of $C(t)$ at the resonance, due to the relaxation of the molecule
by successive photon emissions.

A quite different situation occurs for a non-integer value of the ratio $%
\omega /\Omega $, as shown in Fig. 3 for $m=9,31$. There are revivals of the
initial state at latter times, in the form of an absorption-emission cycle
not completely periodic.

The above results were obtained within first order perturbation theory.
Consequently, the long-time behavior should include corrections due to
multiple photon processes. However, we believe that the essential physics is
contained in the results discussed above. To support this view, in the next
section we analyze the radiation field using the semiclassical
approximation. We predict a masser effect at resonances.

\section{Semiclassical Theory}

\hspace*{4ex} The semiclassical theory is based on two fundamental 
assumptions: i) the cavity radiation field is described by a coherent state, 
which is the most nearly classical state (it minimizes the uncertainty relations); 
ii) the total density matrix is written as a product, whose factors are related to 
spin and photon degrees of freedom. In other words, we assume that spin variables 
are uncorrelated with those of the field (Sargent) \cite{[sargent]}. For a coherent 
state $\left|z\right\rangle$, we have the results
\begin{eqnarray}
a \left|z\right\rangle=z \left|z\right\rangle ,\\
\left\langle z\right|a^{\dagger }=\left\langle z\right|z^*,\\
\left\langle n \right\rangle=\left|z\right|^2 .
\end{eqnarray}
So, the semiclassical approximation can be thought to be obtained by replacing the
photon operators by complex numbers. Chosing the amplitude of the coherent state as real, we get  $a^\dagger,a \rightarrow 
\sqrt{\left\langle n \right\rangle}$. Then the equation of motion for the mean number of photons $\left\langle n \right\rangle$ will be given by: 
\[
\frac{dA}{dt} = -\Gamma \sin(\omega t) [\rho _{12} e^{i r \sin(\Omega
t)}+\rho _{21} e^{-i r \sin(\Omega t)}] \label{eqdadt}
\]
with $A=\sqrt{\left\langle n \right\rangle}$ The density matrix $\rho$ of the spin system satisfies the following set of coupled equations: 
\begin{eqnarray}
\frac{\partial M}{\partial t} = 4 i \Gamma[\rho _{21} e^{-i r \sin(\Omega 
t)}-\rho _{12} e^{i r \sin(\Omega t)}] A \cos(\omega t)  \label{eqMdt} \\
\frac{\partial \rho _{12}}{\partial t} = -2 i \Gamma e^{-i r \sin(\Omega t)}
M A \cos(\omega t)  \label{eqrho12} \\
\frac{\partial \rho _{21}}{\partial t} = 2 i \Gamma e^{i r \sin(\Omega t)} M
A \cos(\omega t)  \label{eqrho21},
\end{eqnarray}
where $M=\rho _{11}-\rho _{22}$. In general the solution must be
accomplished by numerical methods. A simple solution is obtained if one assumes 
$\rho _{12} = \rho_{21}^* = i N$, being $\rho _{12} - \rho _{21} = 2 i N$, \emph{i.e.} the off-diagonal elements depends on a single real parameter. It can be shown that the other possibilty, $\rho _{12} = \rho_{21}= R $ does not lead to a maser effect. For the former case we
have: 
\begin{eqnarray}
\frac{\partial M}{\partial t} = 8 \Gamma N A \cos(r \sin(\Omega t)) 
\cos(\omega t) ,  \label{eqMdt1} \\
\frac{\partial N}{\partial t} = -2 \Gamma M A \cos(r \sin(\Omega t))
\cos(\omega t) ,  \label{eqdndt} \\
\frac{dA}{dt} = 2 \Gamma N \sin(r \sin(\Omega t))\sin(\omega t) .
\label{eqdAdt1}
\end{eqnarray}
An approximate solution near the values $A_0 \approx 0$, $N_0$ and $M_0$ for
the photon field $A$ is easily obtained yielding: 
\begin{equation}  \label{solutionA}
A(t) = A_0 + \Gamma N_0 \mathrm{Re}(F_1-F_2)
\end{equation}
In the case of $\omega = (2 m + 1) \Omega$ the above solution for $A$ grows
linearly with time: 
\begin{equation}  \label{SolutionA2}
A(t) \approx A_0 +2 \Gamma N_0 J_{2 m +1}(r)\left(t-\frac{\sin(2\omega t)}{%
2\omega}\right)
\end{equation}
and consequently, the number of photons increases quadratically with time. In contrast, for
$\omega = (2 m + 2) \Omega$ the approximate solution is nearly constant, with $A(t)\approx A_0$, displaying ripples around this value. This even-odd symmetry breaking at resonance, 
is at variance with the pure quantum case treated in the previous section, where both instances presented similar behaviors at resonance, \emph{i.e.} an increasing of the photon number due to relaxation of the molecular states, no matter if $m$ was even or odd. In any case, for the semiclassical approximation, a maser effect is predicted for $\omega = (2 m + 1) \Omega$.
To illustrate this phenomenon, we show in Figures 4 to 7, 
the photon number as a function of time, for some temperatures. In Fig. 4
and 5, we take the limit $k_B T\rightarrow 0$, and the initial
photon number $n_0\rightarrow 0(A_0\rightarrow 0)$. Other parameter values are $M_0=0$ and $N_0=0.5$. 
In Fig. 4, we have chosen $m=9$ and $r=10$, while for Fig. 5, $m=10$ and $r=12$ 
(the corresponding value of $r$ is always chosen to maximized the Bessel function).
From Fig. 4, it is clear that the overall behavior of the photon number for $m$ odd
is closely related to the solution given by expression (\ref{SolutionA2}), and shown by 
the dashed line, corresponding to a quadratic increase of the number of
photons with time. In contrast, when $m$ is even, there is an oscillatory
behavior, at least for short times, which reminds the quantum revivals.
Figures 6 and 7 refer to examples at finite temperatures for $m$ odd.
For Fig. 6, $T=24$ K, corresponding to the initial value $n_0=k_B T/(\hbar \omega) = 49$ 
for the number of photons, while Fig. 7 is at room temperature ($T=300$K),
corresponding to $n_0 = 625$ photons. Note that the higher the initial
number of photons, the faster the photon number increases with time. The effect 
is paramount in the latter case. Note also the different time scale.

It practice, the photon number will increase until a saturation limit.
The divergent behavior here obtained is due to the fact that losses are not
taken into account. Such losses are provided by photons leaving the
cavity and by excitation of other energy states. In
fact, the real system which we are concerned here, is not as simple 
as a two-level object.  
In addition, the cavity field should include contributions from 
other photon modes, which we 
neglected from the beginning.

\section{Conclusions}

We have developed the quantum dynamics of molecular
nanomagnets, taking into account the quantized photon field of the cavity. 
In first approximation, the magnetic molecules were treated as two-level systems. The temporal
evolution of quantum states were studied in details in this paper,
considering a time-varying magnetic field $B_z(t)=B_0
\cos(\Omega t)$, applied along the quantization direction. The above field produces the necessary population inversion in a periodic way.

The spin system couples with the dipolar component of the cavity field, 
allowing transitions between both molecular states with photon emission and absorption.
The emission of photons is enhanced at the resonant condition, when the ratio between photon
frequency $\omega$ and the applied field frequency $\Omega$ is an integer
number $m=\omega/\Omega$. At resonance, the energy is pumped from the applied magnetic field to the cavity radiation field, inducing a relaxation process depicted in Fig. 1 and 2. This irreversible process can be understood, if one considers that the probability of creating a  photon increases with the number of photons. This striking phenomenon was illustrated in the previous section, via the semiclassical theory.
In turn, the case of non-integer $m$ produces a revival of the initial state, which means that emission and absorption of photons occur in almost periodic sequences, very similar to Rabi periods.

Qualitative insights can be obtained via the semiclassical
approximation, which is more suitable for practical purposes at room
temperatures. Within this theory, the macroscopic state of the cavity field 
is described by the coherent states introduced by Glauber \cite{[glauber]}. 
In this case, a maser effect is obtained at resonance, when $m$ is an odd number. 
The maser effect is enhanced by temperature, when the initial photon number is increased.\\


\newpage
{\large \textbf{Figure Captions}} ~~\newline

Figure 1: Correlation function $C(t)$ as a function of time, considering $%
\Gamma =2\times 10^{9}$ s$^{-1}$, $\omega=2\pi\times 10^{10}$ rad/s, $\Omega
=\omega/m$, $m=8$ and $r=9,64$.\\

Figure 2: Correlation function $C(t)$ with parameters $\Gamma$ and $\omega$ kept the same
as for the previous figure, $m=11$ and $r=11,94$. \\

Figure 3: Correlation function $C(t)$ for a non-integer ratio $\omega/\Omega$. In this
case we have chosen $m=9,31$ and $r=11,45$. \\

Figure 4: Number of photons for the field $A(t)=\sqrt{\left<n\right>}$ in the
semiclassical approximation using $f=\omega/2\pi = 10$ GHz, $\Gamma=1$
GHz, $M_0=0$, $N_0=0.5$, $n_0=0$, $m=9$ and $r=10$.\\

Figure 5: Number of photons for the field $A(t)=\sqrt{\left<n\right>}$ in the
semiclassical approximation using $f=\omega/2\pi = 10$ GHz, $\Gamma=1$
GHz, $M_0=0$, $N_0=0.5$, $n_0=0.25$, $%
m=10$ and $r=12$.\\ 

Figure 6: Number of photons for the field $A(t)=\sqrt{\left<n\right>}$ in the
semiclassical approximation with $f$, $\Gamma$, $M_0=0$ and $N_0=0.5$ kept
the same as the previous case, $n_0=49$, $m=9$ and $r=10$.\\

Figure 7: Number of photons for the field $A(t)=\sqrt{\left<n\right>}$ in the
semiclassical approximation using $n_0=625$, $m=5$ and $%
r=6.5$. Other parameters kept the same as for the previous figures.\\

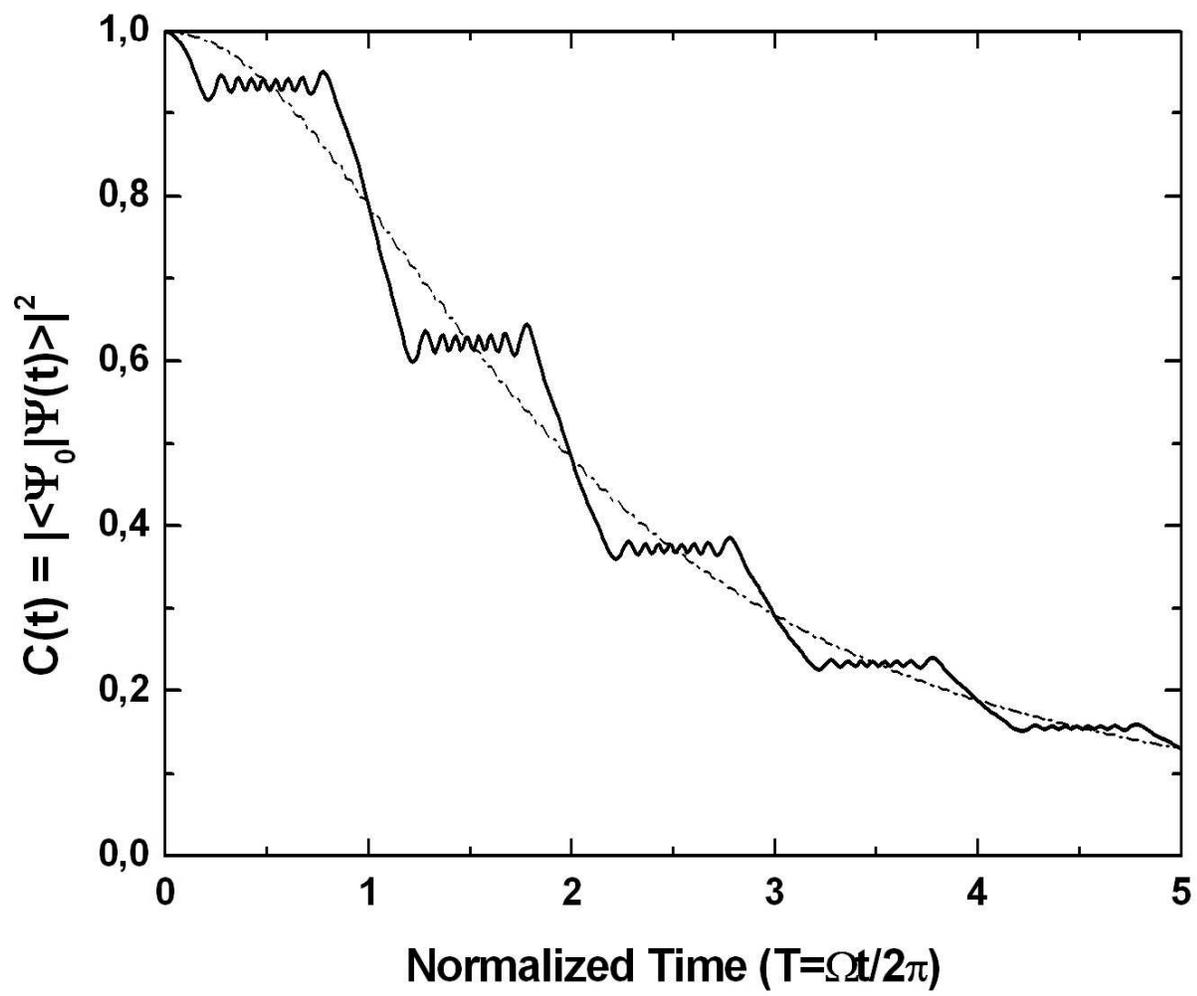
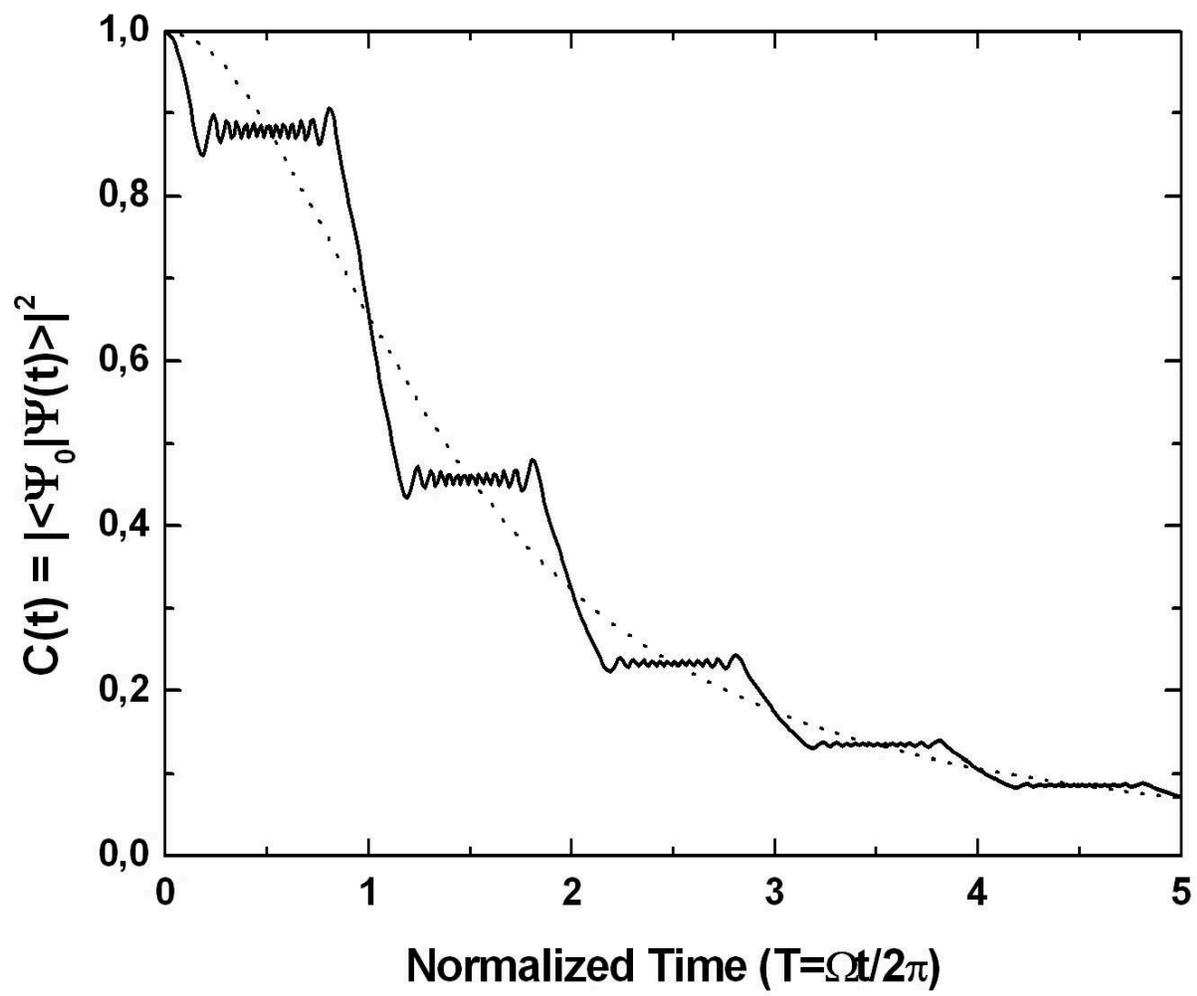
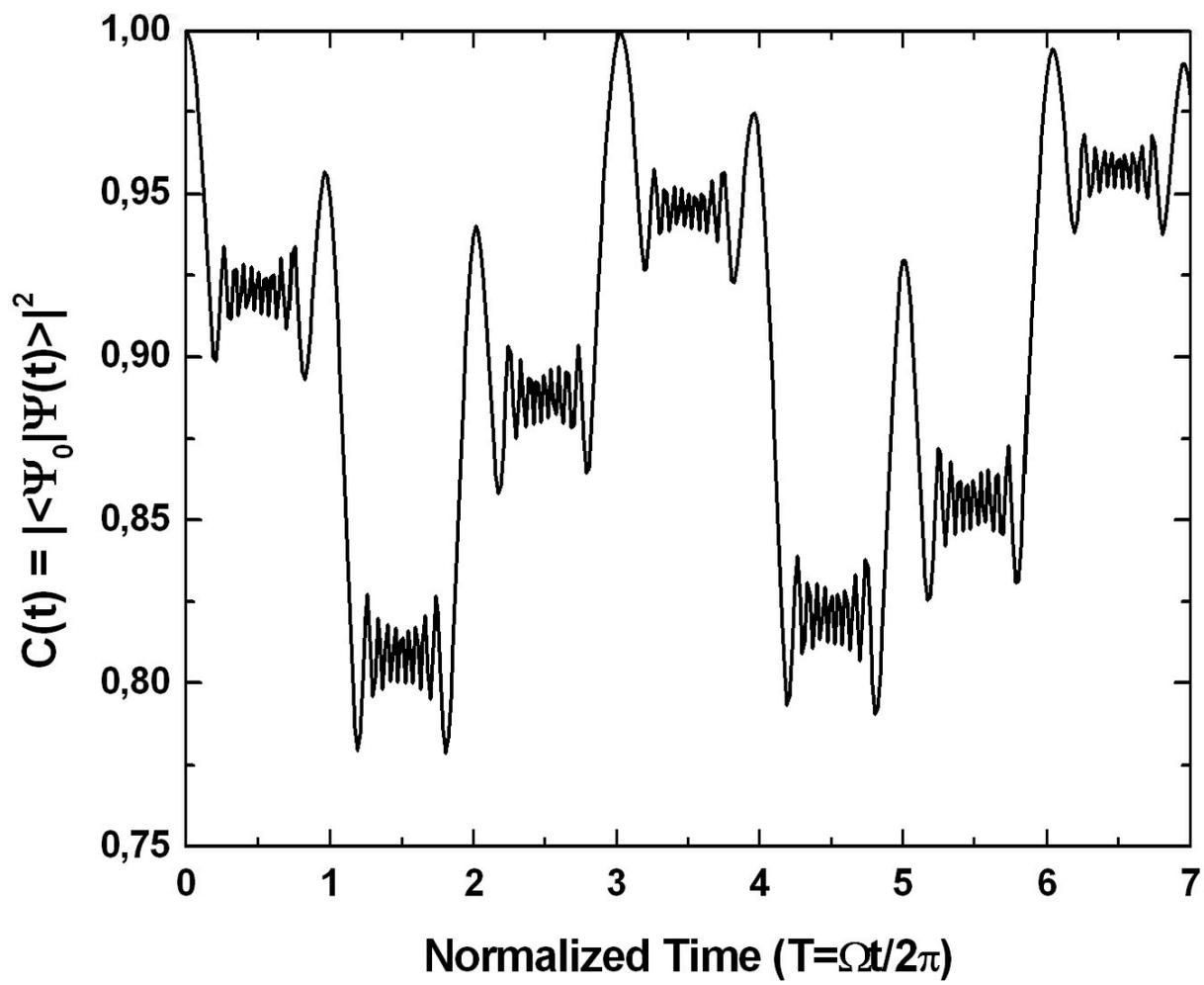
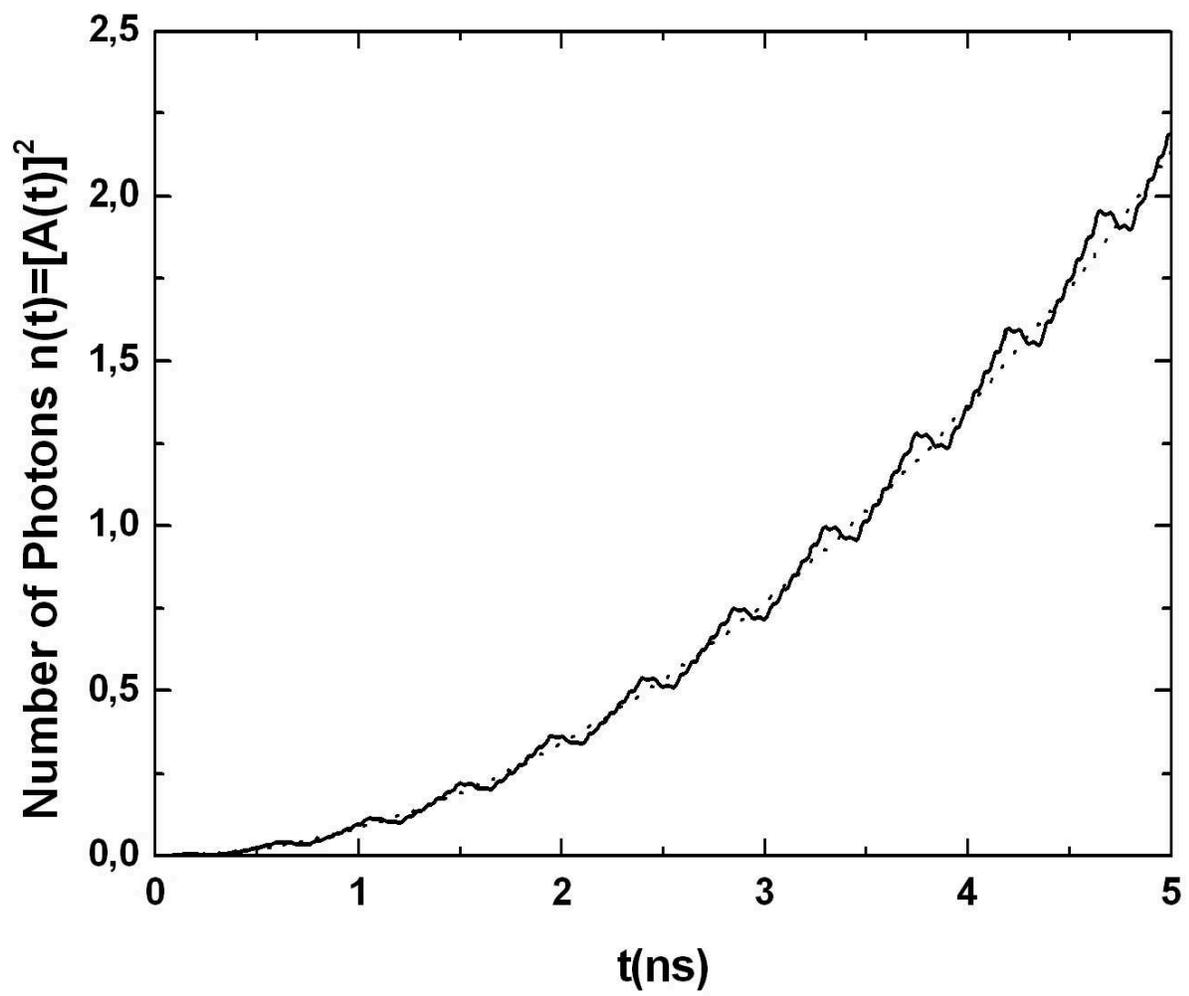
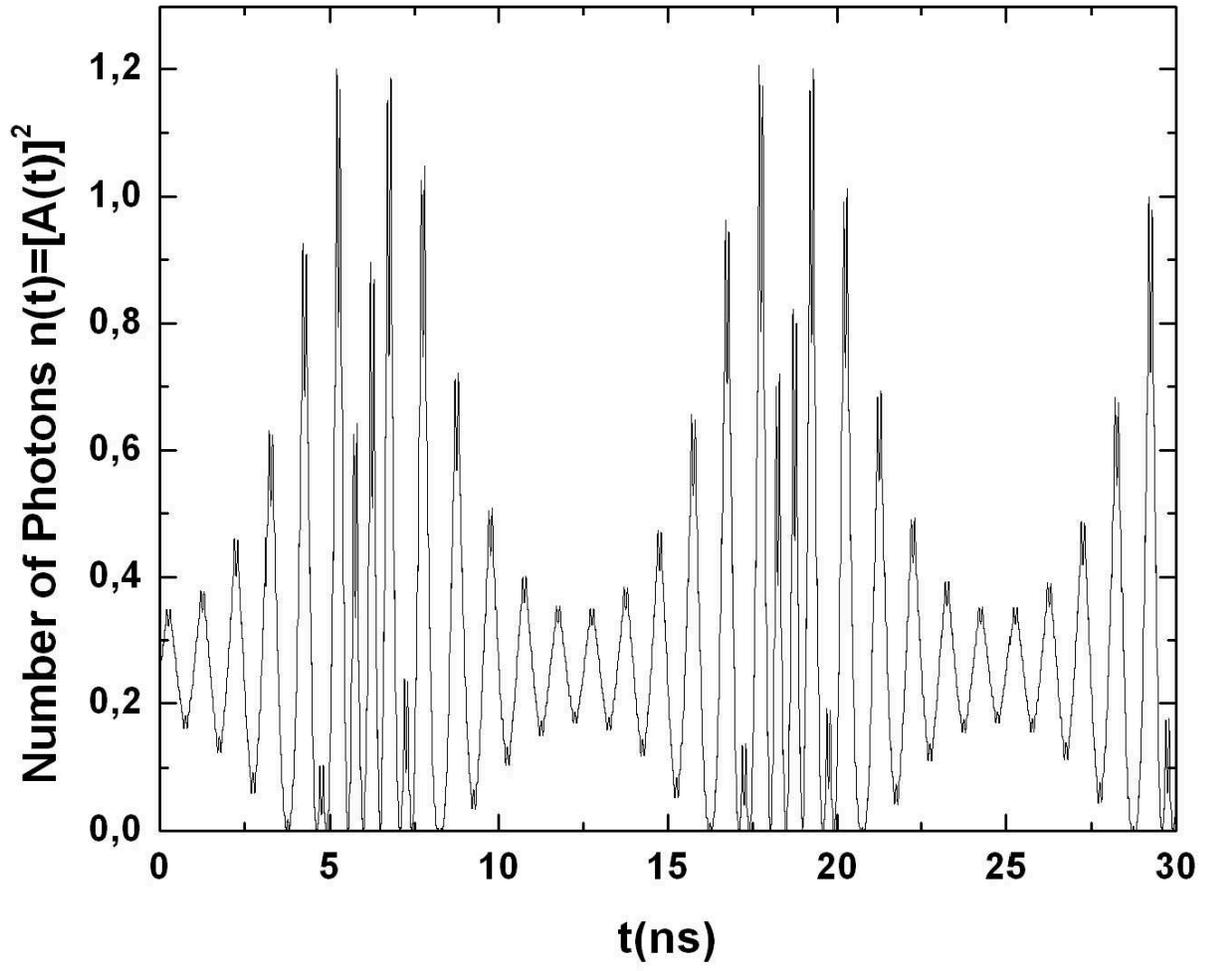
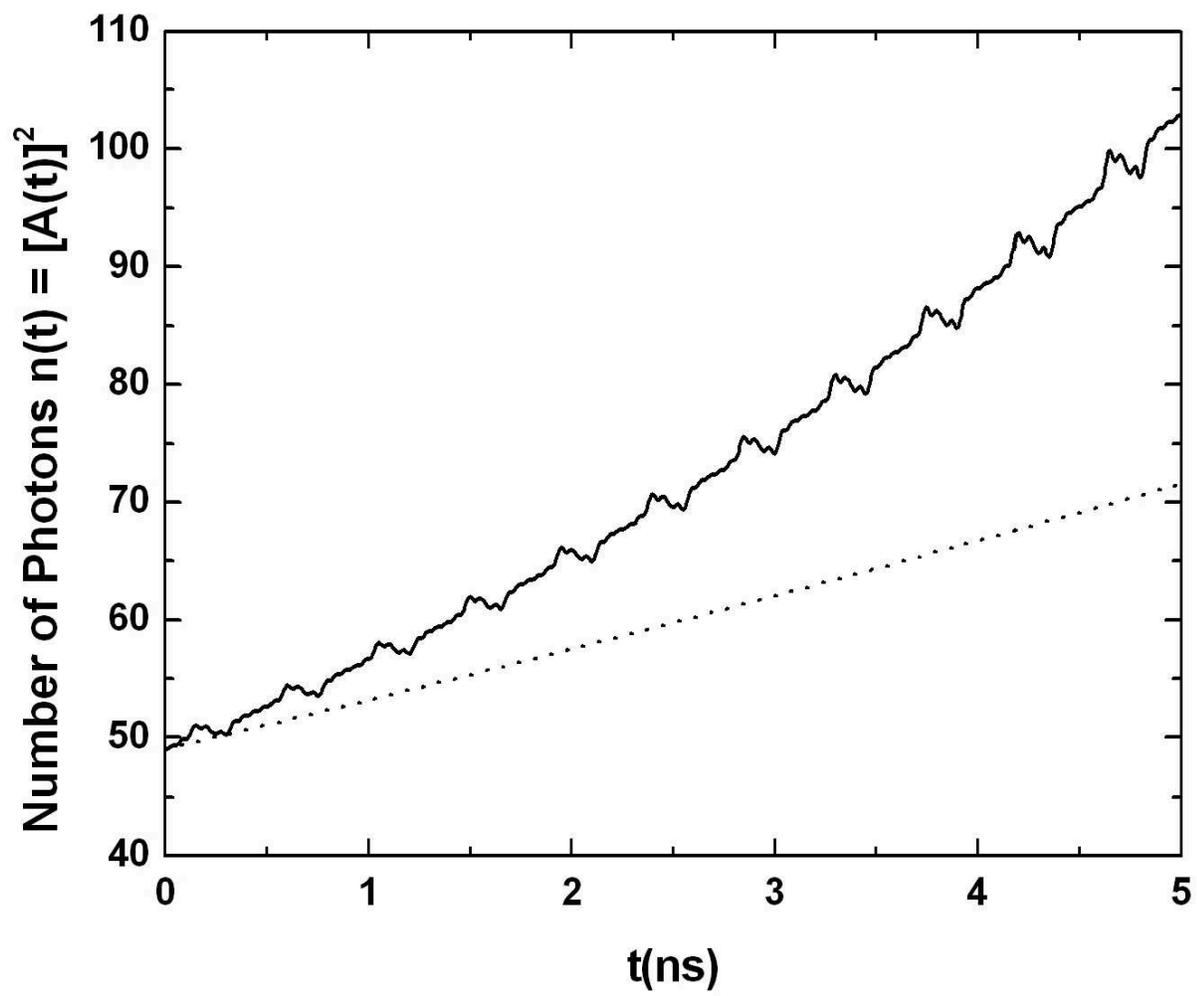
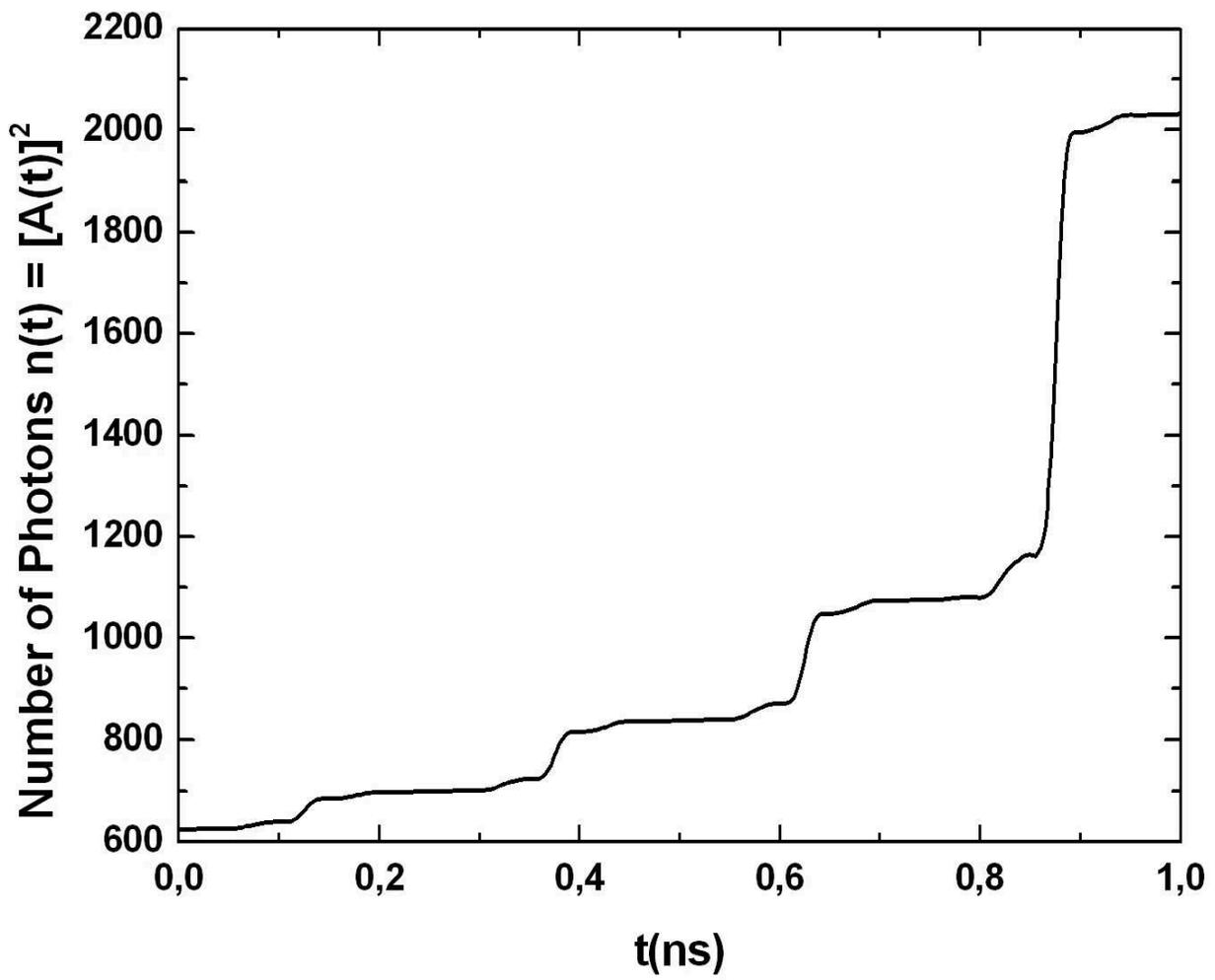
\end{document}